**LETTER • OPEN ACCESS**

# Maize yield and nitrate loss prediction with machine learning algorithms

View the article online for updates and enhancements.

## Recent citations

- Assessing the uncertainty of maize yield without nitrogen fertilization
  Adrian A. Correndo *et al*

- Wild blueberry yield prediction using a combination of computer simulation and machine learning algorithms
  Efrem Yohannes Obsie *et al*

- Impact of extreme weather conditions on European crop production in 2018
  Damien Beillouin *et al*

# Environmental Research Letters

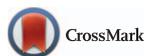

LETTER

OPEN ACCESS

# Maize yield and nitrate loss prediction with machine learning algorithms

Mohsen Shahhosseini[1] , Rafael A Martinez-Feria[2,3] , Guiping Hu[1,4] and Sotirios V Archontoulis[2]

[1] Department of Industrial and Manufacturing Systems Engineering, Iowa State University, Ames, Iowa, United States of America
[2] Department of Agronomy, Iowa State University, Ames, Iowa, United States of America
[3] Department of Earth and Environmental Sciences, Michigan State University, East Lansing, Michigan, United States of America
[4] Author to whom any correspondence should be addressed.

E-mail: gphu@iastate.edu

## Abstract

Pre-growing season prediction of crop production outcomes such as grain yields and nitrogen (N) losses can provide insights to farmers and agronomists to make decisions. Simulation crop models can assist in scenario planning, but their use is limited because of data requirements and long runtimes. Thus, there is a need for more computationally expedient approaches to scale up predictions. We evaluated the potential of four machine learning (ML) algorithms (LASSO Regression, Ridge Regression, random forests, Extreme Gradient Boosting, and their ensembles) as meta-models for a cropping systems simulator (APSIM) to inform future decision support tool development. We asked: (1) How well do ML meta-models predict maize yield and N losses using pre-season information? (2) How many data are needed to train ML algorithms to achieve acceptable predictions? (3) Which input data variables are most important for accurate prediction? And (4) do ensembles of ML meta-models improve prediction? The simulated dataset included more than three million data including genotype, environment and management scenarios. XGBoost was the most accurate ML model in predicting yields with a relative mean square error (RRMSE) of 13.5%, and Random forests most accurately predicted N loss at planting time, with a RRMSE of 54%. ML meta-models reasonably reproduced simulated maize yields using the information available at planting, but not N loss. They also differed in their sensitivities to the size of the training dataset. Across all ML models, yield prediction error decreased by 10%–40% as the training dataset increased from 0.5 to 1.8 million data points, whereas N loss prediction error showed no consistent pattern. ML models also differed in their sensitivities to input variables (weather, soil properties, management, initial conditions), thus depending on the data availability researchers may use a different ML model. Modest prediction improvements resulted from ML ensembles. These results can help accelerate progress in coupling simulation models and ML toward developing dynamic decision support tools for pre-season management.

## 1. Introduction

Traditionally, farmers rely on their experiences and past historical data such as the crop yields and weather to make important decisions to increase short-term profitability and long-term sustainability of their operation (Arbuckle and Rosman 2014). New promising technologies such as machine learning (ML) have emerged over the last years that can potentially aid farmers' decision making (Hoogenboom *et al* 2004, González Sánchez *et al* 2014, Togliatti *et al* 2017, Basso and Liu 2018, Ansarifar and Wang 2018, Moeinizade *et al* 2019). However, the lack of spatial and temporal data that cover a range of production (e.g. yield) and environmental (e.g. N leaching) variables under a range of management inputs (e.g. N-rate, planting date) to efficiently train the ML models is a limitation that needs to be overcome.

Although simulation crop modeling can achieve reasonable prediction accuracy, its application in actual farms is limited because of the substantial amount of expertise and data required for rigorous calibrations (Drummond *et al* 2003, Puntel *et al* 2019). Even with a well-calibrated simulation model, deployment for exploring potential management options under a range of possible weather conditions (i.e. scenario analysis) is often impractical due to long runtimes and data storage constraints. Critically, simulations have to be rerun to incorporate new information as it becomes available, or to extrapolate beyond the set of conditions which were originally simulated.

Meta-models are statistical models trained on more computationally expensive models (e.g. a crop simulator). By replacing a more detailed simulation model, a meta-model can provide faster execution, reduced storage needs, and added flexibility to extrapolate across temporal and spatial scales than a more detailed simulation model (Simpson *et al* 2001). Meta-modeling has been widely implemented for extrapolating hydrological (Fienen *et al* 2015, Nolan *et al* 2018) and biogeochemical (Britz and Leip 2009, Villa-Vialaneix *et al* 2012, Ramanantenasoa *et al* 2019) simulations across regional scales, as well as to streamline sensitivity analyzes for parameterization of crop simulation models (Stanfill *et al* 2015, Pianosi *et al* 2016, Gladish *et al* 2019). To our knowledge, meta-modeling approaches for decision support and forecasting applications in crop production have not been previously evaluated.

The goal of the meta-model is to 'learn' the connections among input and output variables by finding patterns or clusters in the simulated data (Villa-Vialaneix *et al* 2012, Fienen *et al* 2015). The techniques used can range from classical statistical techniques to Machine Learning (ML) algorithms. The latter have enjoyed wide applications in various ecological classification problems and predictive modeling (Rumpf *et al* 2010, Shekoofa *et al* 2014, Crane-Droesch 2018, Karimzadeh and Olafsson 2019, Pham and Olafsson 2019a, 2019b) because of their adeptness to deal with nonlinear relationships, high-order interactions and non-normal data (De'ath and Fabricius 2000). Such methods include regularized regressions (Hoerl and Kennard 1970, Tibshirani 1996, Zou and Hastie 2005), tree-based models (Shekoofa *et al* 2014), Support Vector Machines (Basak *et al* 2007, Karimi *et al* 2008), Neural Networks (Liu *et al* 2001, Crane-Droesch 2018, Khaki and Khalilzadeh 2019, Khaki and Wang 2019) and others.

The central question when developing a meta-model is how well the behavior of the simulation model is reproduced by the selected method. Reasonable predictions (e.g. <20% error) made in real time can be more valuable for quick screening across large geographic regions or high-dimensional factorial spaces than running a more precise but slower simulator (Fienen *et al* 2015). Answering this question requires both examining which predictive methods are best suited to emulate the crop model and determining the requirements of the size and type of the training data. Although the literature is rich on comparisons among the skill of various ML algorithms in predicting agricultural outcomes (e.g. Landau *et al* 2000, Sheehy *et al* 2006, González Sánchez *et al* 2014, Morellos *et al* 2016, Qin *et al* 2018), these are largely based on empirical data. Sensitivity analysis studies have looked at the performance of several approaches to emulate simulated results (Gladish *et al* 2019), but their focus is on approximating the distribution of continuous crop model parameters. Scenario analysis for crop forecasting often uses a combination of both categorical (e.g. crop cultivar, tillage and fertilizer application mode) and continuous (e.g. dates, input amounts, initial conditions) input variables, thus evaluation of ML meta-models with these types of datasets is needed.

In this article, we investigate the potential use of ML algorithms as meta-models for developing more computationally expedient and dynamic decision support systems for crop production. Our goal is to provide a framework for developing robust, fast and dynamic forecasting systems that can provide pre-season (e.g. in April) predictions when information is most needed by farmers, targeting both production (maize yield) and environmental quality (N loss) outcomes. Typically, forecasting systems based on crop models, remote-sensed imagery or surveys provide predictions months after planting (Togliatti *et al* 2017, Basso and Liu 2018). Given the uncertainty in weather, we examined the extent to which the maize yield and N loss target variables can be predicted with ML meta-models trained on pre-season weather information (October to April), initial conditions and management choices. We made use of a large scenario analysis dataset ($n > 3$ million) generated by a well-calibrated simulation model (APSIM) to:

(1) Evaluate the performance of four different ML algorithms as meta-models;

(2) Determine data requirements for achieving acceptable prediction accuracy;

(3) Rank the importance of different data-types on yield and N loss prediction; and

(4) Investigate whether ML ensembles offer better prediction than single algorithms.

## 2. Materials and methods

We first calibrated the APSIM (Agricultural Production Systems SiMulator; Holzworth *et al* 2014) cropping systems model using experimental data from seven locations in the US Midwest with observations of maize yield and N loss in drainage over 5–7 years and few management treatments. Second, we used APSIM to simulate maize yields and N loss responses to a factorial grid of management practices and initial conditions, resulting in more than 3 million simulated data points. Third, we used 88% of the simulated data to develop and train ML algorithms and 12% of the simulated data to evaluate ML algorithms prediction accuracies.

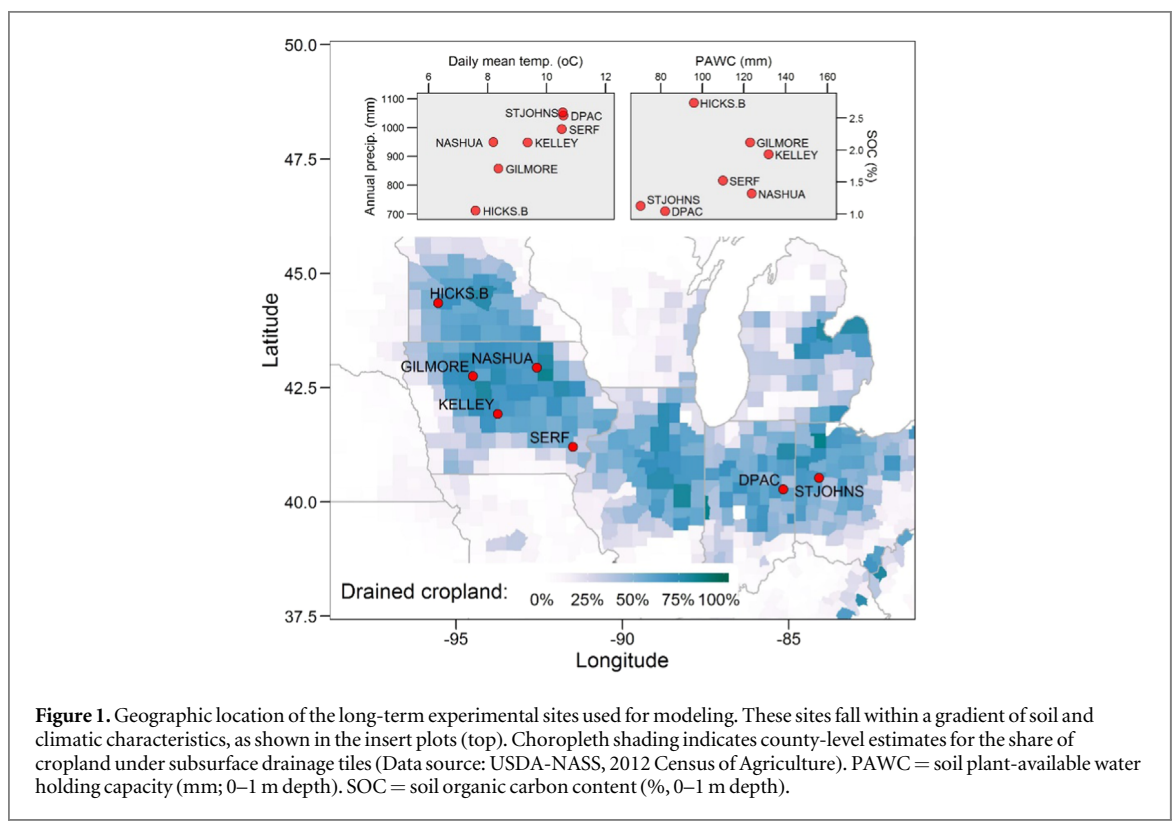

**Figure 1.** Geographic location of the long-term experimental sites used for modeling. These sites fall within a gradient of soil and climatic characteristics, as shown in the insert plots (top). Choropleth shading indicates county-level estimates for the share of cropland under subsurface drainage tiles (Data source: USDA-NASS, 2012 Census of Agriculture). PAWC = soil plant-available water holding capacity (mm; 0–1 m depth). SOC = soil organic carbon content (%, 0–1 m depth).

### 2.1. Experimental locations and data used to train and test APSIM

Figure 1 illustrates the locations of the experimental data used to train and test APSIM. Experimental data for the KELLEY and NASHUA locations have been described in detail in previous modeling studies (Dietzel *et al* 2016, Martinez-Feria *et al* 2018), while data from the remaining locations (DPAC, HICKS.B, GLIMORE, SERF and STJOHNS) were extracted from the Sustainable Corn CAP Research Database (Abendroth *et al* 2017). Soil information for each site was obtained from the SSURGO database (Soil Survey Staff n.d.). Soils in these sites are artificially drained using subsurface drain tubes. Daily weather (1987–2016) for all sites was retrieved from Daymet (Thornton *et al* 2012).

### 2.2. Development of the simulated dataset

We used the calibrated version of the model and performed a factorial simulation experiment to develop a simulated dataset for ML development and analysis. The factorial combinations (see table 1) aimed to characterize the influence of initial conditions, crop management, soil management, and N fertilizer management. Within a factor we included three levels. For instance, for N rate we considered N rate at Maximum Return to Nitrogen (Sawyer *et al* 2006) per location, 30% less and 30% more nitrogen (see table 1). The crop, soil and N management factors were designed to represent levels of practice implementation. The combination of two crops, 26 total factor levels, seven sites, and 34 years (1983–2016), resulted in more than three million simulated scenarios of yield and N loss data. Every model run corresponded to an instance of a full factorial design, so that all possible scenarios were simulated. The simulation process included a re-set on 20th October (average maize harvest time) to decouple the effect of weather-year from the initial conditions. All simulations started on 20th October and ended on 19th October of the next year. Cumulative annual N loss refers to that period. By 'weather-year' we refer to the weather occurring in different years, from 1983 to 2016. We split the weather duration into pre-season (from previous crop harvesting to next crop planting, approximately mid-October to April for our environment) and during the season. For training the ML we used only the pre-season weather as it reflects reality better (the summer weather is unknown when farmers and agronomists make decisions). The pre-season weather's influence is more than planting time. It influences the amount of soil water and depth to water table at planting time and also the amount of soil nitrate held into the soil at planting time. These two variables affect both yield and N leaching predictions.

The developed database included the target variables (yield and N loss) and additional features (explanatory variables) such as data on soil, weather, and management (see supplementary table S2 which is available online at stacks.iop.org/ERL/14/124026/mmedia for a list of variables). To make the ML meta-models more generalizable, the year and location features were replaced by soil characteristics such as texture, soil organic carbon and plant-available water holding capacity and weather data. Weather variables such as temperature and precipitation were integrated for the period 20th October to 10th April (fallow period). The fallow period was divided into five equal periods for each location and weather-year to increase the weather-related features to a total of 35 for ML analysis. No growing season weather data was included in the database because the summer weather is an unknown factor at planting time.

Table 1. Management, cultivar and environmental factors considered in APSIM scenario modeling for creation of the database.

| Category | Variable name | Levels | Levels description |
|---|---|---|---|
| Environment | Weather years | 34 (from 1983 to 2016) | |
| | Locations | KELLEY, DPAC, GILMORE, HICKSB, NASHUA, SERF, SEPAC | |
| Management | Water table | Average | Initialized at the depth of the tile |
| | | Deep | Initialized at 1.1 times tile depth |
| | | Shallow | Initialized at 0.9 times tile depth |
| | Surface residue | Maize, Soybean | Initial residue amount and C:N, derived from spin-up initializations per location |
| | Initial soil N in the profile (0–1 m) | Average | Average soil nitrate per location |
| | | High | 0.5 times the average |
| | | Low | 2 times the average |
| | Sowing time | Temperature rule driven | Plant when 10 d moving average of daily air temperature is > 15 °C |
| | | Delay 10 d | Sow 10 d after |
| | | Delay 20 d | Sow 20 d after |
| | Residue removal | 0% removal | No residue removal |
| | | 33% removal | Remove 33% of standing residue on 20 October |
| | | 66% removal | Remove 66% of standing residue on 20 October |
| | Cover crop | None | No cover crop |
| | | Winterkill | Cover crop planted on 12 October and killed on 1 January |
| | | Overwinter | Cover crop planted on 12 October and killed 10 d before maize sowing |
| | N Strategy | Fall | 100% fertilizer applied as ammonium on 25 November |
| | | Spring | 100% of fertilizer applied as Urea-ammonium nitrate at maize sowing |
| | | Split | 50% of fertilizer applied as Urea-ammonium nitrate at maize sowing and 50% applied 40 d after maize sowing |
| | N rate (kg N/ha) | MRTN | Maximum return to N rate value at each site |
| | | 30% below | 0.7 times the MRTN |
| | | 30% above | 1.3 times the MRTN |
| Genotype | Cultivar | Typical | A calibrated APSIM maize hybrid typical per location |
| | | High YP | Increase radiation-use efficiency by 10% |
| | | High NUE | Decrease N concentration in grains from 1.2% to 1% |

### 2.3. Machine learning meta-models development and testing

Four machine learning meta-models were evaluated to predict simulated maize yield and N loss, including two types of linear regression regularizations (Ridge and LASSO), as well as two tree-based methods (random forests and Extreme Gradient Boosting (XGBoost)). Multiple linear regression was also employed as a baseline to compare performance among techniques. Description for each meta-model technique is provided in the supplementary materials.

Because preliminary analysis indicated that linear regression and regularization meta-models resulted in predictions with high variance (overfitting) when including many explanatory weather features, we trained these meta-models using weather summaries for the whole fallow period instead of the five-period summaries. About 88% of the data available were used for training ML meta-models. Fitting and testing procedures were performed using the R statistical software (version 3.5.0; R core team 2018). We used the *glmnet* package (Friedman *et al* 2010) to fit LASSO and Ridge regression, *ranger* to fit random forests (Wright and Ziegler 2015) and *xgboost* for Extreme Gradient Boosting (Chen and Guestrin 2016).

Partial dependence plots (PDPs) for some of the key input features were used to show the marginal

**Table 2.** Evaluation metrics for various machine learning algorithms applied to predict maize yields and N loss using information up to planting date for the test dataset (2013–2016).

|  | RMSE (kg ha$^{-1}$) | R-RMSE (0%–100%) | $R^2$ (0–1) | RMSE (kg ha$^{-1}$) | R-RMSE (0%–100%) | $R^2$ (0–1) |
|---|---|---|---|---|---|---|
|  | (a) maize yield prediction results | | | (b) N loss prediction results | | |
| Random forest | 1400 | 13.9% | 0.44 | 9.1 | 54.5% | 0.78 |
| XGBoost | 1348 | 13.4% | 0.48 | 16.5 | 98.3% | 0.27 |
| Linear regression | 1489 | 14.8% | 0.37 | 14.4 | 85.6% | 0.44 |
| Ridge regression | 1508 | 14.9% | 0.35 | 14.3 | 85.4% | 0.45 |
| Lasso regression | 1492 | 14.8% | 0.37 | 14.4 | 85.6% | 0.44 |
| Optimal ensemble | 1240 | 12.3% | 0.56 | 8.5 | 51.0% | 0.80 |
| Benchmark ensemble | 1311 | 13.0% | 0.51 | 10.9 | 65.5% | 0.67 |

effect of those selected input variables on the predicted outcomes.

### 2.4. Statistical indices for meta-model performance evaluation

Three measures were used to evaluate the ML meta-model performances. First, the root mean square error (RMSE) that is a measure of difference between predicted and observed values. Second, the Relative Root Mean Square Error (RRMSE) or Normalized RMSE that allows for a direct comparison between different meta-models and meta-model output variables that have different units. Third, the coefficient of determination ($R^2$) that is defined as the proportion of the variance in the response variable that is explained by independent variables. Further, to find the bias of APSIM simulations and ML predictions, the mean bias error (MBE) was calculated. All the equations can be viewed in Archontoulis and Miguez (2014).

### 2.5. Machine learning meta-model validation

We divided the dataset to training (1983–2012) with 2.7 million data and test sets (2013–2016) with 0.4 million data on yield and N loss. The APSIM simulation model is initialized every year on 20 October, and therefore there is no carry over effect from one year to another in terms of soil moisture or nitrogen etc. Because of the possible relationships between different weather-years, a time-wise 5-fold look-forward cross-validation was conducted on the training set to tune the hyperparameters of random forests, XGBoost, ridge regression and LASSO regression (see supplementary figure S3). The difference between this type of cross-validation with the random k-fold cross-validation is that in each fold the built models only predict future observations. In addition, the effect of changing ML hyperparameters on cross-validation and test errors were compared using hyperparameter plots to ensure the cross-validation mode is optimal.

After tuning hyperparameters (ML specific parameter that are calibrated during the learning process) of each of the ML algorithms, a final meta-model for each of them was trained on the whole training set. The hyperparameters that were tuned were: '*mtry*' for random forests, '*nrounds*', '*eta*', and '*gamma*' for XGBoost, and finally '*lambda*' for Ridge and Lasso regression. Note that multiple linear regression does not use hyperparameters. The hold-out test set (2013–2016; 0.4 million data points) was used to compute the prediction error for the test set which is an estimate of the true error for in future (unseen) observations.

### 2.6. Data size requirements and feature importance ranking

To evaluate how sensitive the developed ML meta-models are to the size of the training data, we partitioned the original training set (1983–2012) into subsets containing 5, 10, 15, 20 and 25 continuous weather-years, with 2012 as the last year included in each subset. Each 5 weather-years represents roughly 0.4 million data points. The subsets were used to train the different ML meta-models, which in turn were tested using the hold-out test set (2013–2016).

The importance of data input features (e.g. weather, initial soil N) were quantified for each ML meta-model separately. Permutation importance was used to calculate and compare the importance of input features for each model (Wright and Ziegler 2017; Chen and Guestrin 2016; Friedman et al 2010).

## 3. Results

### 3.1. How well do ML meta-models predict yield and N loss?

Across the entire dataset, maize yield varied from 0 to 15 Mg ha$^{-1}$, with an average of 9,845 kg ha$^{-1}$. Cumulative N loss ranged from 0 to 220 kg N ha$^{-1}$, with an average value of 14.8 kg N ha$^{-1}$. Across all locations, XGBoost and random forests outperformed other ML meta-models in terms of maize yield and N loss prediction in the testing dataset (table 2). Random forests predicted N loss with a higher $R^2$ than maize yield (0.78 versus 0.44, respectively), although RRMSE was lower for yield than N loss (13.9% versus 54.5%, respectively). This is not surprising given that N loss has much higher variation than yields. XGBoost was the best performing ML meta-model in terms of maize yield prediction (RRMSE of 13.4%) (table 2).

The mean bias error (MBE) results indicated that both APSIM simulations and ML predictions were not biased and their values were not consistently higher or lower than real data (see supplementary figures S2, S4 and S5).

PDP for the most accurate models (random forests and XGBoost) showed the marginal effect the most important input features had on the predicted outcome of random forests and XGBoost. For instance, the effect of planting day of year on maize yield predicted values of both these ML models is similar, implying that later planting dates will result in less maize yield. Similarly, it was shown that with the increase in initial soil nitrate, both models predicted more nitrate loss (See supplemental figures S6 and S7 for more details).

In addition, the hyperparameters plot for cross-validation and test errors of the most significant hyperparameters of each ML model are shown in the supplemental figure S8. Based on these results, since the trends for hyperparameters are similar in both CV and test errors, the cross-validation method is finding the distribution of the test set quite good. Also, it was shown that random forests is the least sensitive model to the hyperparameter changes.

Analyzing ML performance within each location revealed different patterns (figure 2). The ML predicted yields and N losses better in some locations than others. In terms of yield prediction, RRMSE ranged from 7% to 29% among locations, while N loss prediction RRMSE ranged from 35% to >100% (figure 2(a)). The highest RRMSE of yield prediction was obtained at one location (GILMORE) and it was consistent for all ML meta-models. In contrast, the highest RRMSE of N loss prediction was at two locations (SERF and HICKS.B), but not consistently among all ML meta-models.

### 3.2. How many data do we need to train ML meta-models?

Different ML meta-models showed distinct sensitivities to the size of the training data (figure 3). In terms of yield prediction, XGBoost was the most sensitive ML meta-model and random forests was the least sensitive ML meta-model to the size of the training dataset. For yield prediction, the maximum RRMSE decreased (thus lower error) when size of the training dataset increased from 0.4 to about 1.6 million data (figure 3). Beyond that point, yield prediction did not benefit much by increasing the size of the training dataset.

In contrast to yield prediction, we did not observe a consistent relationship between ML meta-models RRMSE and size of the data for the N loss prediction (figure 3). Two ML meta-models benefited by increasing the size of the training datasets (regression models) and two were negatively impacted (random forests and XGBoost; figure 3).

### 3.3. Which input data variables are most important for ML prediction?

Different ML meta-models showed different sensitivities to input data variables (figure 4). Considering maize yield predictions, random forests and XGBoost identified weather features, (i.e. temperature and rain) as the most important factors, followed by management and soil properties (i.e. soil organic carbon and plant-available water). The other ML meta-models had a 5-fold lower sensitivity to weather features than random forests and XGBoost in terms of yield prediction. Averaging the sensitivities of all ML meta-models indicated that weather was the most important input feature (figure 4(a)).

Random forests, the best N loss predictor (table 2), indicated weather as the most significant input feature explaining 43% of the total variance and initial conditions as the second most important input feature explaining 38% of the total variance (figure 4). Ridge regression, which is the second best N loss predictor also identified weather more important than others. Linear regression and LASSO similarly found weather followed by management decisions and initial conditions the most important input features. Finally, XGBoost detected initial conditions and weather as the first two most important features (figure 4(b)).

### 3.4. Do ensembles of ML meta-models improve prediction?

Combining different meta-models with different weights resulted in ensembles that improved maize yield and nitrate loss prediction than the best single performing meta-model (table 2). The ML yield ensemble meta-model created an ensemble with four out of our five ML models with a weight of 50% and 33% for random forests and XGBoost respectively. The benchmark ensemble, in which all five ML meta-models received equal weights, performing better than each individual model, stands at the second place after the optimal ensemble. The optimal ensemble resulted in maize yield and N Loss RRMSE of 12.3% and 51%, respectively (table 2).

## 4. Discussion

In this study, we quantified the role of ML algorithms as meta-models for predicting outcomes of high importance to stakeholders such as yield and N loss as early as planting time. We compared ML meta-models for their accuracy (table 1; figure 2), quantified data size requirements (figure 3) and importance of data input variables (figure 4) to inform experimentalists on future data collection protocols and guide modelers on which ML meta-models to choose for predictions. In view of increasing data availability in agriculture and the maturity of analytics from descriptive to prescriptive (National Academy of Sciences, E A M 2019), more ML applications are currently taking place. For example, Ramanantenasoa *et al* (2019) evaluated the performance of various ML based meta-models to emulate the complex process-based models in predicting ammonia emissions produced by agricultural activities and demonstrated the superiority of random forests compared to LASSO regression. Lawes *et al* (2019) used ML and APSIM modeling to predict optimum N rates for wheat, Puntel *et al* (2019) and Qin *et al* (2018) used ML and experimental data to predict optimum N rates to maize, while others are exploring coupling ML and simulation models to develop faster and more flexible tools for impact regional assessments (Fienen *et al* 2015) and simulation model parameterization (Gladish *et al* 2019).

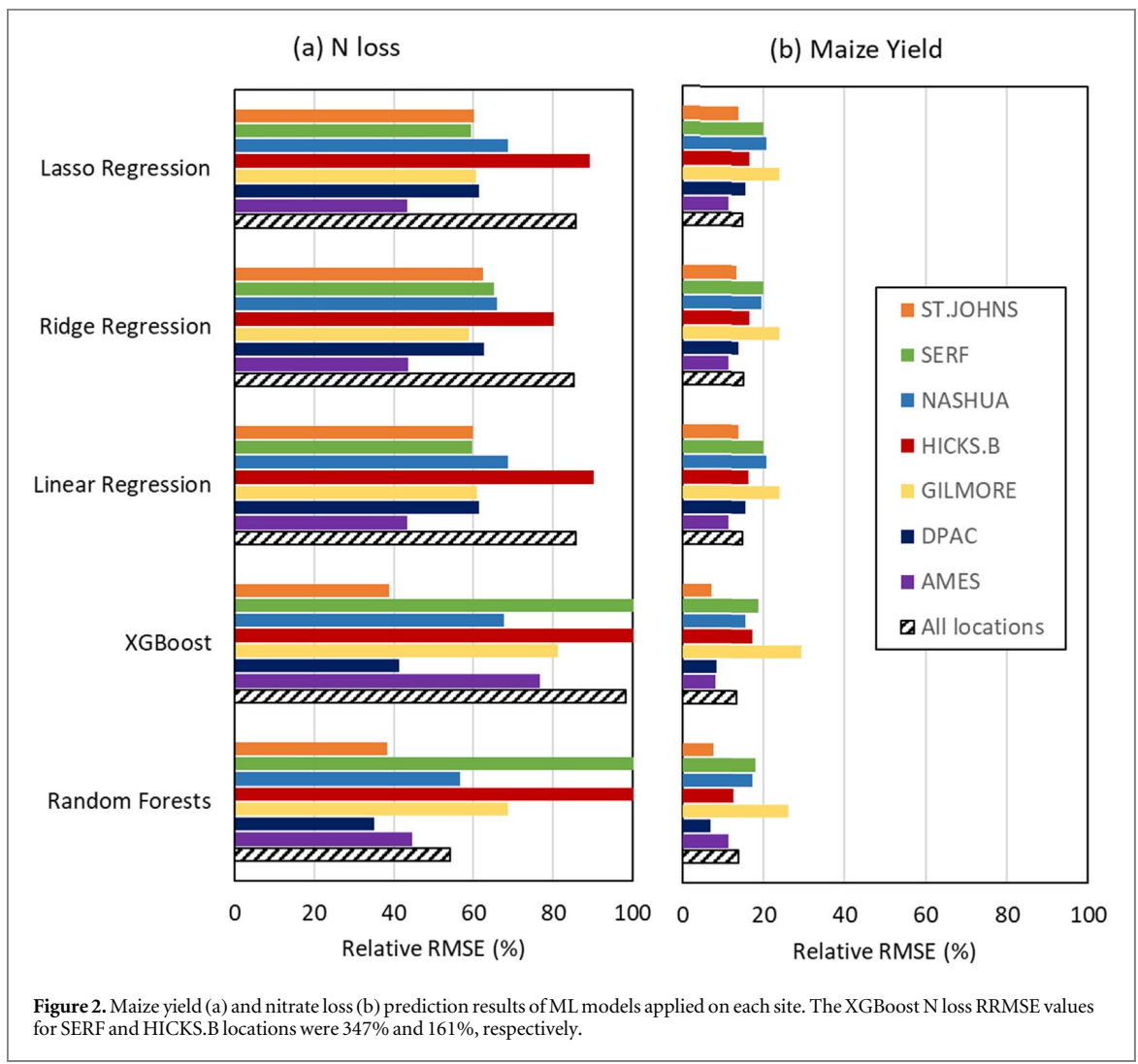

**Figure 2.** Maize yield (a) and nitrate loss (b) prediction results of ML models applied on each site. The XGBoost N loss RRMSE values for SERF and HICKS.B locations were 347% and 161%, respectively.

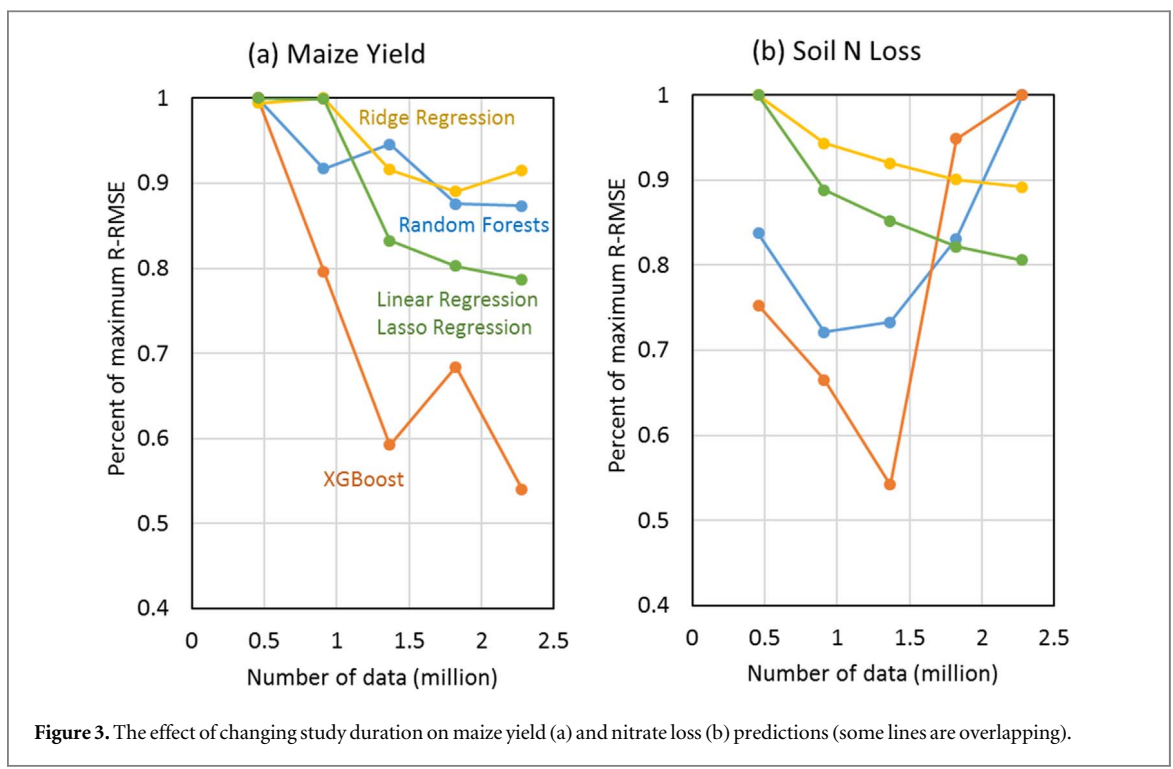

**Figure 3.** The effect of changing study duration on maize yield (a) and nitrate loss (b) predictions (some lines are overlapping).

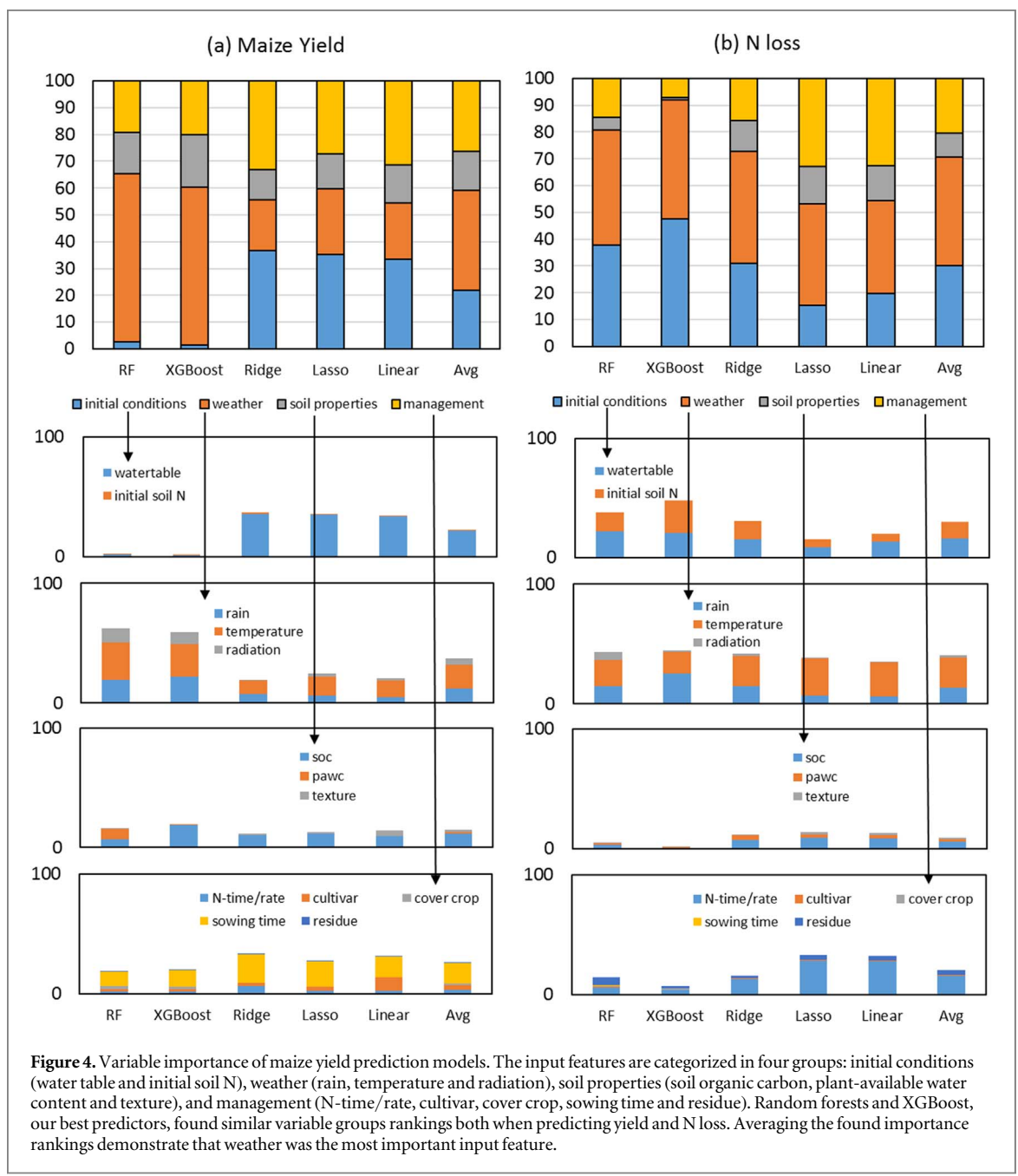

**Figure 4.** Variable importance of maize yield prediction models. The input features are categorized in four groups: initial conditions (water table and initial soil N), weather (rain, temperature and radiation), soil properties (soil organic carbon, plant-available water content and texture), and management (N-time/rate, cultivar, cover crop, sowing time and residue). Random forests and XGBoost, our best predictors, found similar variable groups rankings both when predicting yield and N loss. Averaging the found importance rankings demonstrate that weather was the most important input feature.

The ML algorithms predicted end-of season yield with a RRMSE of 13%–14%, which is comparable to the fit of the simulation model to the field data (figure S3) or even better considering that only information up to planting time was considered in this study (Martinez-Feria *et al* 2018, Puntel *et al* 2019). On the other hand, the RRMSE of cumulative annual N loss (harvest to harvest) was about 4 times higher than for yield and much greater than the error of the simulation model itself, indicating that annual N loss cannot be reliably predicted with information up to planting time. A different approach and most likely more in-season information is needed to reduce the RRMSE in N loss prediction. N loss is very sensitive to extreme rain events, especially those occurring in spring (Iqbal *et al* 2018). So, inclusion of May–June precipitation (a time period that soil N levels are high due to fertilizers and low crop N uptake) into the ML algorithms may improve predictability of this variable.

A common practice to guide management decisions is to use the yield average of previous years. For example, this information is used to estimate yield goals for N fertilizer rate recommendations (Morris *et al* 2018). By using this simple-mean approach, the RMSE associated with yield prediction for the unknown years in our dataset was 1899 kg ha$^{-1}$. Use of ML decreased this error by 20%–29% (table 2), showing its potential added value. Random forests and XGBoost were the best performing ML meta-models for N loss and yield prediction, respectively, among the ML meta-models evaluated (table 2). Random forests is often regarded as a very flexible algorithm, often outperforming other ML techniques across a number of classification and regression problems (Vincenzi *et al* 2011, Mutanga *et al* 2012, Fukuda, *et al* 2013, Jeong *et al* 2016). Improvements in prediction with random forests were much greater in N loss than for yield (table 2), when compared to a multiple linear regression. This could be an indication that N loss has a much greater degree of non-linearity than yields. Despite the modest difference in yield RRMSE, random forests did require less data for training than other ML meta-models (figure 3). Using more than 10 years of weather data did not change much the accuracy of yield prediction. This data requirement is about half compared to all other meta-models.

In addition, different ML algorithms appear to have different sensitivities to the predictor features (figure 4). This may suggest that depending on data input availability for the studied environment, different ML meta-models can be selected. For example, if initial conditions are not of interest, then random forests appears to be the best choice for yield prediction. An interesting result from this analysis is that initial condition becomes very important when ML is to predict N loss and less important when ML is to predict yield. It is important to note that deriving variable importance metrics from ML procedures can produce biased results, depending on the type and the degree of collinearity among the predictors. All of the above needs to be considered when designing protocols for data collection, as well as for future meta-model development.

To evaluate the effect of selecting a subset of most influential input variables on the prediction results, after training random forests model on the entire dataset, first five and ten input variables of the dataset based on random forests feature importance ranking were used to train new random forests models with first five and first ten most important input features. The results showed that random forests could achieve the RRMSE of 14% and 17.3% with only using top ten and top five input features, respectively. Also, RRMSE of random forest for predicting N loss considering top ten and top five features were calculated to be 85.4% and 95.5%, respectively. Therefore, it seems that selecting some of the most influential input features will not result in an improved outcome.

In this study, we faced two major challenges while developing and training the ML meta-models. First, the high variance of the predictions and overfitting due to the dissimilar behavior of different locations in different years meant that the test dataset could not be explained well by the trained ML meta-models. To overcome this challenge, we took the following actions: (1) weather and soil features were added to the dataset to make the meta-models more generalizable, (2) continuous versions of some of the input features were added to the dataset, (3) the fallow season weather information was divided to five equal periods to provide more information for more complex meta-models, and (4) four years (2012–2015; 0.4 million of data) with different weather patterns were chosen as the hold-out test set. The second major difficulty was fine-tuning the hyperparameters of the ML meta-models, given that the behavior of target variables in the different years and different locations were quite dissimilar. We solved this issue with using a 5-fold look-forward cross-validation approach in which we divided the training set to validation and train sets with respect to years and in each fold, the built models only used future observations as the validation set (supplementary figure S3). With this approach, ML meta-models captured the variations of the predictions of all years when those years had been chosen in the validation set.

Interestingly, while both random forests and XGBoost outperformed other meta-models across all locations (table 2), they did not perform well in some locations (figure 2). We attributed this behavior to the existence of heavy outliers (extreme values) in these locations, given the higher propensity of random forests and XGBoost to learn from outliers (Reunanen *et al* 2003). This was one of the reasons why we also examined the possibility of using meta-models. In crop and other modeling applications, model ensembles have been shown to be better predictors of yields than any single simulation model (Asseng *et al* 2013, Wallach *et al* 2018, Kimball *et al* 2019, Shahhosseini *et al* 2019a, 2019b). In this study, we found the same pattern for ML meta-models. In fact, equally weighted ensemble meta-models provided better results than single models when predicting yield (table 2). When different weighting factors were optimized for different ML meta-model combinations for yield prediction, the fit improved compared to the best performing ML meta-model, which in our case was XGBoost (RRMSE of 12.3% versus 13.4%). While ensembles can bring merits of multiple algorithms together, they need the predictions made by single ML meta-models as inputs and also creating ensembles requires more expertise and work. For future work, we recommend evaluating more diverse machine learning algorithms that explain different parts of the variation in the response variable. In doing this, it is expected that the constructed ensembles can show an even better performance. Additionally, optimizing weights of ensembles based on a validation set instead of the test set is another suggestion for future work.

The developed meta-models can offer a feasible option for decision support systems, providing fast and reasonable yield estimates to farmers and agronomists to evaluate decisions. The computation time of APSIM and similar process-based models can be a limitation, especially for large systems that require predictions for exploring a large scenario space in a small amount of time. As an example, comparing runtimes of APSIM and the ML meta-models of 400 randomly chosen scenarios on a personal computer showed that APSIM runtimes for one scenario (5.8 s) were three orders of magnitude greater than all ML models here examined (1–5 ms; figure S9). Already available tools, such as the Soybean Planting Decision Tool released in 2014 (Licht *et al* 2015), have shown their value by helping farmers make decisions especially in challenging environments (e.g. delayed planting in 2019 in Iowa due to an exceptionally wet spring). Our results concur with previous research (Ramanantenasoa *et al* 2019), and highlight the potential of coupling simulation models and ML toward developing dynamic decision support tools for agricultural management.

## 5. Conclusion

We compared four ML algorithms as meta-models to predict simulated maize yield and N loss, trained with available information at planting time. Based on our results, we concluded that simulated yields can be more reasonably emulated than annual cumulative N loss. While the greater performance and lower data requirements of random forests make this the preferred technique among the meta-models evaluated in this analysis, as shown here datasets with heavy outliers might result in high error. Therefore, we recommend that various ML approaches be evaluated for specific datasets when developing meta-models. We found evidence that optimized ML ensembles can substantially outperform the single ML meta-model. This study demonstrates the potential role of meta-models towards developing dynamic recommendation systems for pre-season management decisions.

## Data availability statement

Any data that support the findings of this study are included within the article.

## Acknowledgments

This work is partially supported by project 1011702 from the USDA National Institute of Food and Agriculture and Plant Sciences Institute at Iowa State University.

## ORCID iDs

Mohsen Shahhosseini  https://orcid.org/0000-0001-9588-1143
Rafael A Martinez-Feria  https://orcid.org/0000-0002-4230-5684
Guiping Hu  https://orcid.org/0000-0001-8392-8442
Sotirios V Archontoulis  https://orcid.org/0000-0001-7595-8107